\documentclass[12pt,a4,epsf]{article}
\usepackage{mathrsfs}
\usepackage{graphicx}
\usepackage{latexsym}
\usepackage{epsfig}
\usepackage{bm}
\usepackage{amsmath}
\usepackage{amssymb}
\usepackage{color}
\usepackage{amsthm,ctable}
\usepackage{amsthm}
\usepackage[dvipdfm,CJKbookmarks,bookmarksopen=true,colorlinks=true,
linkcolor=blue,citecolor=blue,pdfstartview=FitH,pdftitle=title,pdfauthor=lixx]{hyperref}
\RequirePackage[mathlines, displaymath]{lineno}      
\usepackage{harvard}
\newtheorem{theo}{\sc Theorem}
\newtheorem{lemm}{\sc Lemma}

\renewcommand{\theequation}{{\arabic{section}.\arabic{equation}}}

\newcommand{\be}{\begin{equation}}
\newcommand{\ee}{\end{equation}}
\newcommand{\bea}{\begin{eqnarray}}
\newcommand{\eea}{\end{eqnarray}}
\newcommand{\beas}{\begin{eqnarray*}}
\newcommand{\eeas}{\end{eqnarray*}}

\makeatletter      
\@addtoreset{equation}{section}
\makeatother       

\long \def\@makecation#1#2{ \vskip 10 pt
\setbox\@tempboxa\hbox{#1:#2} \ifdim \wd\@tempboxa >\hsize
\unhbox\@tempboxa\par \else \hbox
to\hsize{\hfil\box\@temboxa\hfil} \fi}

\def\Xi{X^{(i)}}

\begin{document}


\title{SIMEX estimation for single-index model with covariate measurement error}

\author{ Yiping Yang$^{1}$, Tiejun Tong$^{2}$ and Gaorong Li$^{3}$
\\
\\
{\small $^1$College of Mathematics and Statistics, Chongqing
Technology and Business University,} \\
{\small Chongqing 400067, P. R. China }\\
{\small $^2$Department of Mathematics, Hong Kong Baptist University, Hong Kong}\\
{\small $^3$Beijing Institute for Scientific and Engineering Computing, Beijing University of} \\
{\small Technology, Beijing 100124, P. R. China}
}

\date{}
\maketitle

\begin{quote}
\begin{abstract}
In this paper, we consider the single-index measurement error model with mismeasured covariates in the nonparametric part.
To solve the problem, we develop a simulation-extrapolation (SIMEX) algorithm based on the local linear smoother and the estimating equation.
For the proposed SIMEX estimation, it is not needed to assume the distribution of the unobserved covariate.
We transform the boundary of a unit ball in $\mathbb{R}^p$ to the interior of a unit ball in $\mathbb{R}^{p-1}$ by using the constraint $\|\beta\|=1$.
The proposed SIMEX estimator of the index parameter is shown to be asymptotically normal under some regularity conditions.
We also derive the asymptotic bias and variance of the estimator of the unknown link function.
Finally, the performance of the proposed method is examined by simulation studies and is illustrated by a real data example.

\end{abstract}


\noindent {\footnotesize {\it AMS2000 Subject Classifications}:
primary 62G05, 62G08; secondary 62G20}
\end{quote}

\newpage
\baselineskip=18pt

\section{Introduction}

One major problem in fitting multivariate nonparametric regression models is the ``curse of dimensionality".
To overcome the problem, the single-index model has played an important role in the literature.
In this paper, we consider the single-index model of the form
\begin{eqnarray}
Y=g(\beta^TX)+\varepsilon,\label{SIM}
\end{eqnarray}
where $Y$ is the response variable, $X$ is a $p\times 1$ covariate vector, $g(\cdot)$ is the unknown link function,
$\beta=(\beta_1,\ldots,\beta_p)^T$ is the unknown index parameter, and $\varepsilon$ is a random error with $E(\varepsilon|X)=0$ almost surely.
We further assume the Euclidean norm $\|\beta\|=1$ for the identifiability purpose.
Model (\ref{SIM}) reduces the covariate vector into an index which is a linear combination of covariates,
and hence avoids the ``curse of dimensionality".

Single-index models have been extensively studied in the literature.
See, for example, \citeasnoun{HT1993}, \citeasnoun{HHI1993}, \citeasnoun{CFG1997}, \citeasnoun{XUEZ2006}, \citeasnoun{LZXF2010},
\citeasnoun{LLL2013}, \citeasnoun{LLL2015}, among others.
For estimating the index parameter and the unknown link function,
\citeasnoun{DL1991} developed the sliced inverse regression method.
\citeasnoun{HT1993} proposed the average derivative method to obtain a root-$n$ consistent estimator of the index vector $\beta$.
\citeasnoun{CFG1997} used the local linear method to estimate the unknown parameters and the unknown link function for generalized partially linear single-index models.
\citeasnoun{NT2000} proposed the partial least squares estimator for single-index models.
\citeasnoun{XUEZ2006} and \citeasnoun{ZX2006} proposed the bias-corrected empirical likelihood method to construct the confidence intervals or regions of the parameters of interest. \citeasnoun{LLL2010} proposed the semiparametrically efficient profile least-squares estimators of regression coefficients for partially linear single-index models.
\citeasnoun{ZH2010} extended the generalized likelihood ratio test to the single-index model.
\citeasnoun{CWZ2011} introduced the estimating function method to study the single-index models.
\citeasnoun{PX2012} and \citeasnoun{YXL2014} investigated the single-index random effects models with longitudinal data.
\citeasnoun{LPDT2014} constructed the simultaneous confidence bands for the nonparametric link function in single-index models.

In this paper, we are interested in estimating the index parameter $\beta$ and the unknown link function $g(\cdot)$ in model (\ref{SIM})
when the covariate vector $X$ is measured with error.
We assume an additive measurement error model as
\begin{eqnarray}
W=X+U,\label{EV}
\end{eqnarray}
where $W$ is the observed surrogate, $U$ follows $N(0,\Sigma_u)$ and is independent of $(X,Y)$.
When $U$ is zero, there is no measurement error.
For simplicity, we consider only the case where the measurement error covariance matrix $\Sigma_{u}$ is known.
Otherwise, $\Sigma_{u}$ need to be first estimated, e.g., by the replication experiments method in \citeasnoun{CRSC2006}.
We refer to the models characterized by (\ref{SIM}) and (\ref{EV}) as the single-index measurement error model.

The measurement error models arise frequently in practice and are attracting attention in medical and statistical research.
For example, covariates such as the blood pressure \cite{CRSC2006} and the CD4 count \cite{LC2000,LH2009} are often subject to measurement error. For a class of
generalized linear measurement error models,
\citeasnoun{Stefanski1989} and \citeasnoun{Nakamura1990} used a method of moment identities to construct the corrected score functions,    \citeasnoun{YLT2015}  further developed the corrected empirical likelihood method. \citeasnoun{CS1994} developed the SIMEX method to correct the effect estimates in the presence of additive measurement error.
\citeasnoun{CLKS1996} further investigated the asymptotic distribution of the SIMEX estimator.
Since then, the SIMEX method has become a standard tool for correcting the biases induced by measurement error in covariates for many complex models.
\citeasnoun{CMR1999} and \citeasnoun{DH2008} applied the SIMEX technique to  local polynomial nonparametric regression and  spline-based regression.
\citeasnoun{LR2005} applied the SIMEX technique to the generalized partially linear models with the linear covariate being measured with additive error.
Other interesting works in SIMEX include, for example, \citeasnoun{CZ2003}, \citeasnoun{MC2006}, \citeasnoun{AC2009}, \citeasnoun{ML2010}, \citeasnoun{MY2011},  \citeasnoun{SM2014}, \citeasnoun{ZZZ2014}, \citeasnoun{CL2015}, and \citeasnoun{WW2015}.

Note that the aforementioned SIMEX methods may not be able to handle the multivariate nonparametric measurement error regression models owing to the ``curse of dimensionality".
In view of this, \citeasnoun{LW2005} considered the partially linear single-index measurement error models with the linear part containing the measurement error,
where they applied the correction for attenuation approach to obtain the efficient estimators of the parameters of interest.
Their method, however, is not applicable for the occurrence with measurement errors in the nonparametric part.
This motivates us to develop a new SIMEX method to solve this problem.
Specifically, we combine the SIMEX method, the local linear approximation method, and the estimating equation to handle the single-index measurement error model.
Our method has several desirable features.
First, our proposed method can deal with multivariate nonparametric measurement error regression and avoids ``curse of dimensionality" by introducing the index parameter.
Second, we use the SIMEX technique to construct the efficient estimation and reduce the bias of the estimator, and do not assume the distribution of the unobservable $X$.
Third, to obtain the efficient estimator of $\beta$, we regard the constraint $\|\beta\|=1$ as a piece of prior information and adopt the ``delete-one-component" method.

The remainder of the paper  is organized as follows.
In Section 2, we develop the SIMEX algorithm to obtain the estimators of the index parameter and the unknown link function, and investigate their asymptotic properties.
In Section 3, we present and compare the results from simulation studies and also apply the proposed method to a real data example for illustration.
Some concluding remarks are given in Section 4, and the proofs of the main results are given in the Appendix.

\section{Main Results}

\subsection{Methodology}
To conduct efficient estimation for $\beta$ in the presence of covariate measurement error, \citeasnoun{CS1994} introduced the SIMEX algorithm.
The SIMEX algorithm consists of the simulation step, the estimation step, and extrapolation steps.
It aims to add additional variability to the observed $W$ in order to establish the trend between the measurement error induced bias and the variance of induced measurement error,
and then extrapolate this trend back to the case without measurement error \cite{CRSC2006}.
In this section, we use the SIMEX algorithm, the local linear smoother and the estimating equation to estimate $\beta$ and $g(\cdot)$.
First, we estimate $g(\cdot)$ as a function of $\beta$ by using  the local linear smoother.
We then estimate the parametric part based on the estimating equation. The proposed algorithm is described as follows.

\vskip 12pt
\noindent
\textbf{(I) Simulation step}

 For each $i=1,\ldots,n$, we generate a sequence of variables
 \begin{eqnarray*}
W_{ib}(\lambda)=W_i+(\lambda\Sigma_u)^{1/ 2}U_{ib}, ~~~b=1,\ldots, B,
\end{eqnarray*}
where $U_{ib}\sim N(0,I_p)$, $I_p$ is a $p\times p$ identity matrix, $B$ is a given integer,
and $\lambda\in\Lambda=\{\lambda_1,\lambda_2,\ldots,\lambda_M\}$ is the grid of $\lambda$ in the extrapolation step.  We set the range from 0 to 2.

\vskip 12pt
\noindent
\textbf{(II) Estimation step}

Suppose that $g(\cdot)$ has a continuous second derivative. For $t$ in a small neighborhood of $t_0$, $g(t)$ can be approximated as
$g(t)\approx g(t_0)+g'(t_0)(t-t_0)\equiv a+b(t-t_0)$.
With the simulated $W_{ib}(\lambda)$, we first estimate $g(t_0)$ as a function of $\beta$ by a local linear smoother, denoted by $\hat{g}(\beta, \lambda; t_0)$, in Step 1.
We then propose a new estimator of $\beta(\lambda)$ in Steps 2 and 3, denoted by $\hat{\beta}(\lambda)$. The specific procedure is as follows.

\vskip 6pt
\textbf{Step 1.} For each fixed $t_0$ and $\beta$, $\hat{g}(\beta, \lambda; t_0)$ and $\hat{g}'(\beta,\lambda; t_0)$ are estimated by minimizing
\begin{equation}\label{LS}
\sum\limits_{i=1}^n\left\{Y_{i}-a-b[\beta^TW_{ib}(\lambda)-t_0]\right\}^2K_h(\beta^TW_{ib}(\lambda)-t_0),
\end{equation}
with respect to $a$ and $b$, where $K_h(\cdot)=h^{-1}K(\cdot/h)$, $K(\cdot)$ is a kernel function with $h$ the bandwidth.
Let $\hat{a}$ and $\hat{b}$ be the solutions to problem (\ref{LS}). Then,
$\hat{g}(\beta, \lambda; t_0)=\hat{a}$ and
$\hat{g}'(\beta, \lambda; t_0)=\hat{b}$. Let
\begin{align*}
M_{ni}(\beta,\lambda;t_0)             &= U_{ni}(\beta,\lambda;t_0)\Big/\sum\limits_{j=1}^nU_{nj}(\beta,\lambda;t_0), \\
\widetilde{M}_{ni}(\beta,\lambda;t_0) &= \widetilde{U}_{ni}(\beta,\lambda;t_0)\Big/\sum\limits_{j=1}^nU_{nj}(\beta,\lambda;t_0),
\end{align*}
where
$U_{ni}(\beta,\lambda;t_0)=K_h(\beta^TW_{ib}(\lambda)-t_0)\{S_{n,2}(\beta,\lambda;t_0)-[\beta^TW_{ib}(\lambda)-t_0]S_{n,1}(\beta,\lambda;t_0)\}$,
$\widetilde{U}_{ni}(\beta,\lambda;t_0)=K_h(\beta^TW_{ib}(\lambda)-t_0)\{[\beta^TW_{ib}(\lambda)-t]S_{n,0}(\beta,\lambda;t_0)-S_{n,1}(\beta,\lambda;t_0)\}$,
and $S_{n,l}(\beta,\lambda;t_0)=\dfrac{1}{n}\sum\limits_{i=1}^n(\beta^TW_{ib}(\lambda)-t_0)^lK_h(\beta^TW_{ib}(\lambda)-t_0)$ for $l=0,1,2$.
Simple calculation yields
\begin{align}
\hat{g}(\beta, \lambda; t_0)  &= \sum\limits_{i=1}^nM_{ni}(\beta, \lambda; t_0)Y_{i}, \label{hatg} \\
\hat{g}'(\beta, \lambda; t_0) &= \sum\limits_{i=1}^n\widetilde{M}_{ni}(\beta, \lambda; t_0)Y_{i}. \label{hatg1}
\end{align}

\citeasnoun{CXZ2010} showed that the coverage rate of the estimator of $g'(t)$ is slower than that of $g(t)$ if the same bandwidth is used.
Because of this, we have suggested another bandwidth $h_1$ to control the variability in the estimator of $g'(t)$.
We use $h_1$ to replace $h$ in $\hat{g}'(\beta, \lambda; t_0)$ and write it as $\hat{g}'_{h_1}(\beta, \lambda; t_0)$.

\vskip 6pt
\textbf{Step 2.} To estimate $\beta$, we use the ``delete-one-component" method in \citeasnoun{ZX2006} to transform the boundary of a unit ball in $\mathbb{R}^p$ to the interior of a unit ball in $\mathbb{R}^{p-1}$.
Let $\beta^{(r)}=(\beta_1,\ldots,\beta_{r-1},\beta_{r+1},\ldots,\beta_p)$ be a $(p-1)$ dimensional vector deleting the $r$th component $\beta_r$.
Without loss of generality, we assume there is a positive component $\beta_r$; otherwise, we may consider $\beta_r=-(1-\|\beta^{(r)}\|^2)^{1/2}$.
Let
$$\beta=(\beta_1,\ldots, \beta_{r-1}, (1-\|\beta^{(r)}\|)^2)^{1/2},\beta_{r+1},\ldots,\beta_p)^T.$$
Note that $\beta^{(r)}$ satisfies the constraint $\|\beta^{(r)}\|<1$.
We conclude that $\beta$ is infinitely differentiable in a neighborhood of $\beta^{(r)}$ and the Jacobian matrix is $J_{\beta^{(r)}}=(\gamma_1,\ldots,\gamma_p)^T$,
where $\gamma_s (1\leq s \leq p, s\neq r)$ is a $(p-1)$ dimensional vector with the $s$th component being 1, and $\gamma_r=-(1-\|\beta^{(r)}\|^2)^{-\frac{1}{2}}\beta^{(r)}$.
Given the estimators  $\hat{g}(\beta, \lambda; t_0)$ and $\hat{g}'_{h_1}(\beta, \lambda; t_0)$  in (\ref{hatg}) and
(\ref{hatg1}), respectively,
an estimator of $\beta^{(r)}$, $\hat{\beta}_b^{(r)}(\lambda)$, is obtained by solving the following equation:
\begin{eqnarray}\label{EF}Q_{nb}(\beta^{(r)},\lambda)=
{1\over n}\sum\limits_{i=1}^n\hat{\eta}_{ib}(\beta^{(r)},\lambda)=0,
\end{eqnarray}
where
\begin{align*}
\hat{\eta}_{ib}(\beta^{(r)},\lambda) &= [Y_i-\hat{g}(\beta,\lambda; \beta^TW_{ib}(\lambda))]\hat{g}'_{h_1}(\beta,\lambda; \beta^TW_{ib}(\lambda))J^T_{\beta^{(r)}}W_{ib}(\lambda), \\
\beta^TW_{ib}(\lambda)               &= {\beta^{(r)T}}W_{ib}^{(r)}(\lambda)+(1-\|\beta^{(r)}\|^2)^{1/2}W_{ib,r}(\lambda), \\
W_{ib}^{(r)}(\lambda)                &= (W_{ib,1}(\lambda),\ldots,W_{ib,(r-1)}(\lambda),W_{ib,(r+1)}(\lambda),\ldots,W_{ib,p}(\lambda))^T.
\end{align*}
Next, we can obtain an estimator of $\beta$, say $\hat{\beta}_b(\lambda)$, by implementing the Fisher's method of scoring version of the Newton-Raphson algorithm to solve the estimating equation (\ref{EF}).  We summarize the iterative algorithm in what follows.

(1) Choose the initial values for $\beta$, denoted by $\widetilde{\beta}_b(\lambda)$, where $b=1,\ldots,B$.

(2)  Update $\widetilde{\beta}_b(\lambda)$ with $\widetilde{\beta}_b(\lambda)=\hat{\beta}_b^*(\lambda)/\|\hat{\beta}_b^*(\lambda)\|$ by
$$\hat{\beta}_b^*(\lambda)=\widetilde{\beta}_b(\lambda)+J_{\widetilde{\beta}^{(r)}}B_{nb}^{-1}(\widetilde{\beta}^{(r)},\lambda)Q_{nb}(\widetilde{\beta}^{(r)},\lambda),$$
where $B_{nb}({\beta}^{(r)},\lambda)=\dfrac{1}{n}\sum\limits_{i=1}^nJ^T_{\beta^{(r)}}W_{ib}(\lambda)\hat{g}'^2_{h_1}(\beta,\lambda; \beta^TW_{ib}(\lambda))W_{ib}^T(\lambda)J_{\beta^{(r)}}$.

(3) Repeat Step (2) until convergence.

In the iterative algorithm, the initial values of $\beta$, $\beta_{{\rm {int}}}$, with norm 1 is obtained by fitting a linear model.

\vskip 6pt
{\emph{Remark 1.}  Similar to \citeasnoun{CWZ2011}, we discuss the solution of the estimating equation. In fact, the solution of the estimating equation $Q_{nb}(\beta^{(r)},\lambda)$ is just the least-squares estimator of $\beta^{(r)}$. The least-squares objective function is defined by
$${G}(\beta^{(r)},\lambda)=\sum\limits_{i=1}^n\{Y_i-\hat{g}(\beta,\lambda; \beta^TW_{ib}(\lambda))\}^2.$$
The minimum of the objective function ${G}(\beta^{(r)},\lambda)$ with respect to $\beta^{(r)}$ is the solution of the estimating equation $Q_{nb}(\beta^{(r)},\lambda)$ because the estimating equation $Q_{nb}(\beta^{(r)},\lambda)$ is the gradient vector of ${G}(\beta^{(r)},\lambda)$.
Note that $\{\|\beta^{(r)}\|<1\}$ is an open, connected subset of $\mathbb{R}^{p-1}$. By the regularity condition (C2), we known that the least-squares objective function ${G}(\beta^{(r)},\lambda)$ is twice continuously differentiable on $\{\|\beta^{(r)}\|<1\}$ such that the global minimum of
${G}(\beta^{(r)},\lambda)$ can be achieved at some point.  By some simple calculations, we have
$$\frac{1}{n}\frac{\partial^2G(\beta^{(r)},\lambda)}{\partial\beta^{(r)}\beta^{(r)T} }=-\frac{\partial Q_{nb}(\beta^{(r)},\lambda)}{\partial\beta^{(r)}}=\mathcal{A}(\beta(\lambda),\lambda)+o_p(1),$$
where $\mathcal{A}(\beta(\lambda),\lambda)$ is a positive definite matrix for $\lambda\in\Lambda$ defined in Condition (C6).  Then, the Hessian matrix $\dfrac{1}{n}\dfrac{\partial^2G(\beta^{(r)},\lambda)}{\partial\beta^{(r)}\beta^{(r)T} }$ is positive definite for all values of
$\beta^{(r)}$ and $\lambda\in\Lambda$. Hence, the estimating equation (\ref{EF}) has a unique solution.

\vskip 12pt
\textbf{Step 3.} With the estimated values $\hat{\beta}_b(\lambda)$ over $b=1,\ldots, B$, we average them and obtain the final estimate of $\beta$ as
$$\hat{\beta}(\lambda)={1\over B}\sum\limits_{b=1}^B\hat{\beta}_b(\lambda).$$

\vskip 6pt
\noindent
\textbf{(III) Extrapolation step}

For the extrapolant function, we consider the widely used quadratic function $\mathcal{G}(\lambda,\Psi)=\psi_1+\psi_2\lambda+\psi_3\lambda^2$ with $\Psi=(\psi_1,\psi_2,\psi_3)^T$ \cite{LC2000,LR2005}.
We fit a regression model of $\{\hat{\beta}(\lambda),\lambda\in\Lambda\}$ on $\{\lambda\in\Lambda\}$ based on $\mathcal{G}(\lambda,\Gamma)$,
and denote $\hat{\Gamma}$ as the estimated value of $\Gamma$.
The SIMEX estimator of $\beta$ is then defined as $\hat{\beta}_{{\rm SIMEX}}=\mathcal{G}(-1,\hat{\Gamma})$.
When $\lambda$ shrinks to 0, the SIMEX estimator reduces to the naive estimator, $\hat{\beta}_{{\rm Naive}}=\mathcal{G}(0,\hat{\Gamma})$, that neglects the measurement error with a direct replacement of $X$ by $W$.

The SIMEX estimator, $\hat{g}_{\rm SIMEX}(t_0)$, is obtained in the same way.
 $\beta$ in Step 1 of the estimation step is replaced by $\hat{\beta}_{\rm SIMEX}$ and the estimator $\hat{g}_b(\lambda; t_0)$ is obtained with the bandwidth $h_2$. $\hat{g}_b(\lambda; t_0)$  over $b=1,\ldots, B$ is averaged, then $\hat{g}(\lambda; t_0)$ is obtained by
$$\hat{g}(\lambda; t_0)={1\over B}\sum\limits_{b=1}^B\hat{g}_b(\lambda; t_0).$$
The extrapolation step results in $\hat{\mathbb{A}}$, which minimizes $\sum_{\lambda\in\Lambda}\{\hat{g}(\lambda;t_0)-\mathcal{G}(\lambda;\mathbb{A})\}^2$ with respect to $\mathbb{A}$.
The SIMEX estimator of $\hat{g}_{\rm SIMEX}(t_0)$
is given by
$$\hat{g}_{\rm SIMEX}(t_0)=\mathcal{G}(-1,\hat{\mathbb{A}}).$$

\subsection{Asymptotic properties}

To investigate the asymptotic properties of the estimators for the index parameter and the link function, we first present some regularity conditions.
\begin{enumerate}
\item[(C1)] The density function, $f(t)$, of $\beta^T X$ is bounded away from zero. It also satisfies the Lipschitz condition of order 1 on $\mathcal{T}= \{t = \beta^T x : x\in A\}$, where $A$ is the bounded support set of $X$.

\item[(C2)] $g(\cdot)$ has a continuous second derivative on $\mathcal{T}$.

\item[(C3)] The kernel $K(\cdot)$ is a bounded and symmetric density function with a bounded support satisfying  the Lipschitz condition of order 1 and $\int_{-\infty}^{\infty} u^2K(u)du\neq 0$.

\item[(C4)] $\sup\limits_x E(\varepsilon^2|X = x) < \infty$ and $\sup\limits_x E(\varepsilon^4|X = x) < \infty$.

\item[(C5)] $nh^2/(\log n)^2\rightarrow \infty$, $nh^4 \log n\rightarrow 0$, $nhh_1^3/(\log n)^2\rightarrow \infty$, and $\lim \sup\limits_{n\rightarrow\infty}nh_1^5<\infty$.

\item[(C6)] $\mathcal{A}(\beta(\lambda),\lambda)$ is a positive definite matrix for $\lambda\in\Lambda$, where
$$\mathcal{A}(\beta(\lambda),\lambda)=E\Big\{\Big[g'\Big(\lambda; \beta^T(\lambda) W_{ib}(\lambda)\Big)\Big]^2J_{\beta^{(r)}(\lambda)}^T{\widetilde{W}}_{ib}(\lambda){\widetilde{W}}^T_{ib}(\lambda)J_{\beta^{(r)}(\lambda)}\Big\}$$ with
${\widetilde{W}}_{ib}(\lambda)=
{{W}}_{ib}(\lambda)-E[{{W}}_{ib}(\lambda)| \beta^T(\lambda) W_{ib}(\lambda)]$.

\item[(C7)] The extrapolant function is theoretically exact.
\end{enumerate}

{\emph{Remark 2.}
Condition (C1) ensures that the the density function of $\beta^TX$ is positive.
Condition (C2) is the standard condition in smoothness.
Condition (C3) is the common assumption for the second-order kernels.
Condition (C4) is a necessary condition for deriving the asymptotic normality for the proposed estimator.
Condition (C5) specifies some mild condition for the choice of bandwidth.
Finally, Condition (C6) ensures that there is asymptotic variance for the estimator $\hat{\beta}_{\rm SIMEX}$,
and Condition (C7) is the common assumption for the SIMEX method.

To derive the theoretical results, we introduce some new definitions and notations.
For the given $\Lambda=\{\lambda_1,\ldots,\lambda_M\}$, let $\hat{\beta}(\Lambda)$ be the vector of estimators $(\hat{\beta}(\lambda_1),\ldots,\hat{\beta}(\lambda_M))$,
denoted by ${\rm vec}\{\hat{\beta}(\lambda),\lambda\in\Lambda\}$.
Let also $\mathbf{\Gamma}=(\Gamma_1^T,\ldots,\Gamma_p^T)^T$, where $\Gamma_j$ is the parameter vector estimated in the extrapolation step for the $j$th component of $\hat{\beta}(\lambda)$ for $j=1,\ldots,p$.
We define $\mathcal{G}(\Lambda,\mathbf{\Gamma})={\rm vec}\{\mathcal{G}(\lambda_m, \Gamma_j), j=1,\ldots,p, m=1,\ldots,M\}$,
${\rm Res}(\mathbf{\Gamma})=\hat{\beta}(\Lambda)-\mathcal{G}(\Lambda,\mathbf{\Gamma})$, $s^T(\mathbf{\Gamma})=\{\partial/\partial (\mathbf{\Gamma})^T\}{\rm Res}(\mathbf{\Gamma})$,
$D(\mathbf{\Gamma})=s(\mathbf{\Gamma})s^T(\mathbf{\Gamma})$,
$$\eta_{iB}(\beta(\lambda),\lambda)=\displaystyle {1\over B}\sum\limits_{b=1}^B\Big[Y_i-g\Big(\lambda; \beta^T(\lambda) W_{ib}(\lambda)\Big)\Big]g'\Big(\lambda; \beta^T(\lambda) W_{ib}(\lambda)\Big)J_{\beta^{(r)}(\lambda)}^T{\widetilde{W}}_{ib}(\lambda),$$
$$\Psi_{iB}\Big\{\beta(\Lambda),\Lambda\Big\}={\rm vec}\{\eta_{iB}(\beta(\lambda),\lambda),\lambda\in\Lambda\},$$
$$\mathcal{J}\Big\{\beta(\Lambda),\Lambda\Big\}={\rm diag}\{J_{\beta^{(r)}(\lambda)},\lambda\in\Lambda\},$$
$$\mathcal{A}_{11}\Big\{\beta(\Lambda),\Lambda\Big\}={\rm diag}\{\mathcal{A}(\beta(\lambda),\lambda),\lambda\in\Lambda\}$$
and
$$
\Sigma=\mathcal{J}\Big\{\beta(\Lambda),\Lambda\Big\}\mathcal{A}^{-1}_{11}\Big\{\beta(\Lambda),\Lambda\Big\}C_{11}\Big\{\beta(\Lambda),\Lambda\Big\}
\Big\{\mathcal{A}^{-1}_{11}\Big\{\beta(\Lambda),\Lambda\Big\}\Big\}^T\mathcal{J}^T\Big\{\beta(\Lambda),\Lambda\Big\}
$$
with
$$C_{11}\Big\{\beta(\Lambda),\Lambda\Big\}={\rm cov}\Big[\Psi_{iB}\Big\{\beta(\Lambda),\Lambda\Big\}\Big].$$

\begin{theo}\label{AN}
Suppose that the regularity conditions $(C1)$--$(C7)$ hold. Then, as $n\rightarrow \infty$,  we have
$$\sqrt{n}(\hat{\beta}_{\rm SIMEX}-\beta)\stackrel{\mathcal {L}}{\longrightarrow}N\{0,\mathcal{G}_{\mathbf{\Gamma}}(-1,\mathbf{\Gamma})\Sigma(\mathbf{\Gamma})
\{\mathcal{G}_{\mathbf{\Gamma}}(-1,\mathbf{\Gamma})\}^T\},$$
where $\stackrel{\mathcal {L}}{\longrightarrow}$ denotes the
convergence in distribution, $\mathcal{G}_{\mathbf{\Gamma}}(\lambda,\mathbf{\Gamma})=\{\partial/\partial(\mathbf{\Gamma})^T\}\mathcal{G}(\lambda,\mathbf{\Gamma})$,
$\Sigma(\mathbf{\Gamma})=D^{-1}(\mathbf{\Gamma})s(\mathbf{\Gamma})\Sigma s^T(\mathbf{\Gamma})D^{-1}(\mathbf{\Gamma})$.
\end{theo}

Theorem \ref{AN} indicates that $\hat{\beta}_{\rm SIMEX}$ is a root-$n$ consistent estimator.
Its asymptotic distribution is similar to that of the parametric estimator of $\beta$ without measurement error,
whereas the asymptotic covariance matrix of the resulting estimator is more complicated.

Let $f_0(\cdot)$ be the density function of $\beta^TW$, $\mu_l=\int t^lK(t)dt$ and $\nu_l=\int K^l(t)dt$ for $l=1,2$.
Define $$\gamma(\lambda,\mathbb{A})=\{\partial/\partial(\mathbb{A})\}\mathcal{G}(\lambda,\mathbb{A}),$$
$$C(\Lambda,\mathbb{A})=\gamma^T(-1,\mathbb{A})\Big\{\sum\limits_{\lambda\in\Lambda}\gamma(\lambda,\mathbb{A})\gamma^T(\lambda,\mathbb{A})\Big\}^{-1},$$
and $D=E_q\gamma(\lambda,\mathbb{A})\gamma^T(\lambda,\mathbb{A})E_q$,
where $E_q$ is the $q\times q$ matrix of all elements being zero except for the first element being one and $q$ is the dimension of $\mathbb{A}$.

\begin{theo}\label{ANG}
Suppose that the regularity conditions $(C1)$--$(C7)$ hold, and assume that $nh_2^5=O(1)$. Then, as $n\rightarrow\infty$ and $B\rightarrow\infty$,
the SIMEX estimator $\hat{g}_{\rm SIMEX}(t_0)$ is asymptotically equivalent to an estimator whose bias and variance are given respectively by
$$C(\Lambda,\mathbb{A})\sum\limits_{\lambda\in\Lambda}{1\over 2}h_2^2 \mu_2g''(\lambda;t_0)\gamma(\lambda,\mathbb{A})$$
and
$$[nh_2f_0(t_0)]^{-1}\nu_2 {\rm var}(Y|\beta^TW=t_0)C(\Lambda,\mathbb{A})DC^T(\Lambda,\mathbb{A}),$$
where $g(\lambda; t)=E(Y|\beta^TW_b(\lambda)=t).$
\end{theo}

Theorem \ref{ANG} implies that the $\hat{\beta}_{\rm SIMEX}$ does not affect the estimator of $\hat{g}_{\rm SIMEX}(t_0)$ because $\hat{\beta}_{\rm SIMEX}$
is root-$n$ consistent. As pointed out in \citeasnoun{CMR1999}, the variance of $\hat{g}_{\rm SIMEX}(t_0)$ is asymptotically the same as if the measurement error was ignored, but multiplied by a factor, $C(\Lambda,\mathbb{A})DC^T(\Lambda,\mathbb{A})$, which is independent of the regression function.

\section{Numerical studies }

\subsection{Simulation study}
In this section, we evaluate the finite sample performance of the proposed method via simulation studies. Consider the following model
\begin{eqnarray*}
\left\{
  \begin{array}{ll}
   Y_i=-2(\beta^TX_i-1)^2+1+\varepsilon_i,  \\
    W_i=X_i+U_i,\qquad i=1,\ldots,n,
  \end{array}
\right.
\end{eqnarray*}
where $\beta=(\beta_1,\beta_2)^T=(\sqrt{3}/3,\sqrt{6}/3)^T$, $X_i$ is a two-dimensional vector with independent $N(0,1)$ components, the
error $\varepsilon_i$ is generated from $N(0,0.2^2)$, $Y_i$ is generated according to the model, $U_i$ is generated from $N(0, \hbox{diag}(\sigma_u^2,0))$. We
take $\sigma_u = 0.2, 0.4$ and 0.6 to represent different levels of measurement errors. In simulation study, we
compare the naive estimates (Naive) that ignore measurement errors and the SIMEX estimates with quadratic extrapolation function. The sizes of the
samples are $n = 50, 100$ and 150. For each setting, we simulate 500 times to assess the performance.
Using the SIMEX algorithm, we take $\lambda=0,0.2,\ldots,2$ and $B=50$.
We use the Epanechnikov kernel  $K(u)=0.75(1-u^2)_{+}$.
As pointed out in \citeasnoun{LW2005}, the computation is quite expensive for the SIMEX method.
In view of this, we apply a ``rule of thumb" to select the bandwidths, which is the same in spirit as the selection method in \citeasnoun{AC2009}.
Specifically, the bandwidths $h$, $h_1$ and $h_2$ are taken to be $cn^{-1/4}(\log n)^{-1/2}$, $cn^{-1/5}$ and $cn^{-1/5}$,
where $c$ is the standard deviation of $\beta_{\rm {int}}^TW$. To explained the rationality of the ``rule of thumb" (RT), we compare with the results of simulations by using the cross-validation (CV) method to select the bandwidths. We apply the same bandwidths for each $\lambda$ and $b$ since it is time consuming for the CV method. The CV statistic is given by
$${\rm{CV}}(h)=\frac{1}{n}\sum\limits_{i=1}^n\{Y_i-\hat{g}_{[i]}(\hat{\beta}^T_{[i]}X_i)\}^2,$$
where $\hat{g}_{[i]}(\cdot)$ and $\hat{\beta}_{[i]}$ are the SIMEX estimators of ${g}(\cdot)$ and $\beta$ which are computed with all
of the samples but the $i$th subject deleted. The $h_{\rm opt}$ is obtained by minimizing ${\rm{CV}}(h)$. It can be shown $h_{\rm opt} = Cn^{-1/5}$ for a constant $C>0$. Therefore, we use the bandwidths
$$h=h_{\rm opt} n^{-1/20}(\log n)^{-1/2}, ~ h_1=h_{\rm opt},~ h_2=h_{\rm opt}.$$

To evaluate the performance of the bandwidth selection for the CV method, we first plot the ${\rm{CV}}(h)$ versus the bandwidth $h$. The simulation result is shown in Figure \ref{CVh} with  $n=100$ and $\sigma_\mu=0.4$ for one run, and other cases are similar. Figure \ref{CVh} shows the relationship
of ${\rm{CV} }(h)$ versus $h$ with $h$ ranging from [0.1, 1]. From Figure \ref{CVh}, we can see that the ${\rm{CV}}(h)$ function is convex, and reaches the minimum value when $h$  is around 0.35.

Table \ref{htable} summarizes the biases and standard deviations (SD) of the parameter $\beta$ obtained by the SIMEX and naive estimators with the two different  bandwidth selections. From Table \ref{htable},  the results of the SIMEX and naive estimators made by different bandwidths have little difference. Hence, to reduce the calculation time, we use the ``rule of thumb" to select the bandwidths in the real data analysis.

\begin{figure}[htbp!]
    \centering
    \includegraphics[width=8cm,height=8cm]{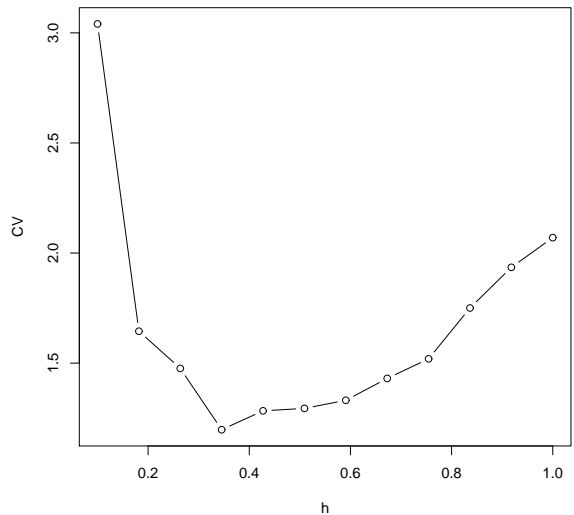}
\caption{{ \it Plot of the ${\rm{CV}}(h)$ versus the bandwidth $h$ with $n=100$ and $\sigma_\mu=0.4$. }}
   \label{CVh}
\end{figure}

\begin{table}[h]
 \caption{\label{htable} {\rm  The biases and standard deviations (SD) of the parameters $\beta_1$ and $\beta_2$
obtained by the SIMEX and naive estimators. }}
\centering \tabcolsep 0.05cm
\begin{tabular}{ cccccccc }
\hline
&&&\multicolumn{2}{c}{SIMEX}&& \multicolumn{2}{c}{Naive}     \\
\cline{4-5}\cline{7-8}
~~~~ &~~~~&~~~~&$\beta_1$&$\beta_2$&&$\beta_1$&$\beta_2$~~
\\\cline{4-5}\cline{7-8}
~~~~$n$~~~~&~$h$~ &~$\sigma_u$~&Bias(SD)&Bias(SD)&&Bias(SD)&Bias(SD)~\\
 \hline
 ~~~50~~~&&0.2&$-0.0084(0.0520)$&$0.0078(0.0377)$&&$-0.0177(0.0291)$&$0.0146(0.0203)$\\
 &$h_{\rm RT}$&0.4&$-0.0405(0.0875)$&$0.0171(0.0638)$&&$-0.0764(0.0537)$&$0.0546(0.0388)$\\
  &&0.6&$-0.0508(0.1253)$&$0.0342(0.0821)$&&$-0.1207(0.0680)$&$0.0700(0.0330)$\\
\cline{2-8}
&   &0.2&$-0.0094(0.0426)$&$0.0031(0.0389)$&&$-0.0182(0.0296)$&$0.0101(0.0206)$\\
&$h_{\rm CV}$&0.4&$-0.0398(0.0867)$&$0.0205(0.0702)$&&$-0.0795(0.0365)$&$0.0508(0.0262)$\\
 &  &0.6& $-0.0548(0.1254)$&$0.0300(0.0845)$&&$-0.1157(0.0710)$&$0.0707(0.0311)$\\
 \hline
100&&0.2&$-0.0083(0.0384)$&$0.0074(0.0321)$&&$-0.0126(0.0203)$&$0.0084(0.0142)$\\
 &$h_{\rm RT}$&0.4&$-0.0381(0.0581)$&$0.0158(0.0334)$&&$-0.0761(0.0397)$&$0.0434(0.0224)$\\
  &&0.6&$-0.0394(0.0719)$&$0.0206(0.0567)$&&$-0.1154(0.0383)$&$0.0632(0.0210)$\\
\cline{2-8}
&   &0.2&$-0.0078(0.0456)$&$0.0076(0.0332)$&&$-0.0119(0.0244)$&$0.0115(0.0185)$\\
&$h_{\rm CV}$&0.4&$-0.0375(0.0521)$&$0.0165(0.0349)$&&$-0.0705(0.0365)$&$0.0454(0.0228)$\\
 &  &0.6&$-0.0391(0.0679)$&$0.0236(0.0449)$&&$-0.1048(0.0379)$&$0.0638(0.0209)$  \\
 \hline
 150&&0.2&$-0.0077(0.0203)$&$0.0050(0.0141)$&&$-0.0187(0.0136)$&$0.0127(0.0093)$\\
 &$h_{\rm RT}$&0.4&$-0.0178(0.0283)$&$0.0117(0.0193)$&&$-0.0497(0.0279)$&$0.0324(0.0177)$\\
  &&0.6&$-0.0279(0.0599)$&$0.0163(0.0394)$&&$-0.1088(0.0315)$&$0.0563(0.0171)$ \\
\cline{2-8}
&   &0.2&$-0.0079(0.0279)$&$0.0044(0.0115)$&&$-0.0181(0.0179)$&$0.0122(0.0119)$\\
&$h_{\rm CV}$&0.4&$-0.0201(0.0294)$&$0.0089(0.0206)$&&$-0.0411(0.0267)$&$0.0383(0.0196)$\\
 &  &0.6& $-0.0252(0.0583)$&$0.0205(0.0371)$&&$-0.0973(0.0357)$&$0.0599(0.0196)$\\
 \hline
\end{tabular}%
\end{table}

Next, we compare the naive estimators and the SIMEX estimators.   From Table \ref{htable}, we can see that the SIMEX estimates of
$\beta_1$ and $\beta_2$ have smaller biases than the naive estimates. However, the standard deviations  based on the SIMEX estimates are larger than those based on the naive estimates. We can also see that the bias and SD decrease as $n$ increases and the estimators depend on the measurement error.

The performance of the estimator for the link function $g(t)$ is discussed by 500 replications. The estimator $\hat{g}(t)$ is
$\hat{g}(t)=\dfrac{1}{500}\sum\limits_{m=1}^{500}\hat{g}_m(t)$. To assess the estimator $\hat{g}(t)$, we use the
root mean squared error (RMSE), which is given by
$$\hbox{RMSE}=\left[n_{\rm grid}^{-1}\sum\limits_{k=1}^{n_{\rm grid}}\{\hat{g}(t_k)-g(t_k)\}^2\right]^{1/2},$$
where $n_{\rm grid}$ is the number of grid points, and $\{t_k, k=1,2,\ldots, n_{\rm grid}\}$ are equidistant grid points. In the simulation study, we take $n_{\rm grid}=15$. The estimated link function and the boxplot for the 500 RMSEs are given in Figure \ref{fighatg}. From Figure \ref{fighatg} (a), we see that the SIMEX estimated curve is closer to the real link function curve than the naive estimated curve. Figure
\ref{fighatg} (b) shows that the RMSEs of the SIMEX and naive estimators for the link function are not large, but
the RMSEs of the SIMEX estimator are slightly larger than the naive estimator.

Note that the SD and RMSE based on the SIMEX estimators are larger than the naive estimators for the parameter $\beta$ and the link function
$g(\cdot)$, respectively. This can be intuitively illustrated with the linear model. Consider the linear model $Y=\beta_0+\beta_x x+\epsilon$, where $E(\epsilon)=0$ and ${\rm Var}(\epsilon)=\sigma_\epsilon^2$. If replacing $x$ with $W+\sqrt{\lambda}\sigma_e e_b$, where $e_b\sim N(0,1)$ and $W=x+e$ with $e$ have mean 0 and variance $\sigma_e^2$, then $\hat{\beta}_{x}(b,\lambda)$ has the asymptotic variance $\{\sigma_\epsilon^2/[\sigma_x^2+(1+\lambda)\sigma_e^2]\}$. If $\lambda=-1$, then ${\beta}_{x}(b,-1)$ is identical to the true parameter, with  the asymptotic variance $\sigma_\epsilon^2/\sigma_x^2$. If $\lambda=0$, ${\beta}_{x}(b,0)$ is just the naive estimator, with the asymptotic variance $\sigma_\epsilon^2/(\sigma_x^2+\sigma_e^2)$.
 Hence, it can be seen easily that the SD or RMSE of the naive estimators is smaller than that of the SIMEX estimators.

\begin{figure}[htbp!]
    \centering
    \includegraphics[width=6cm,height=6cm]{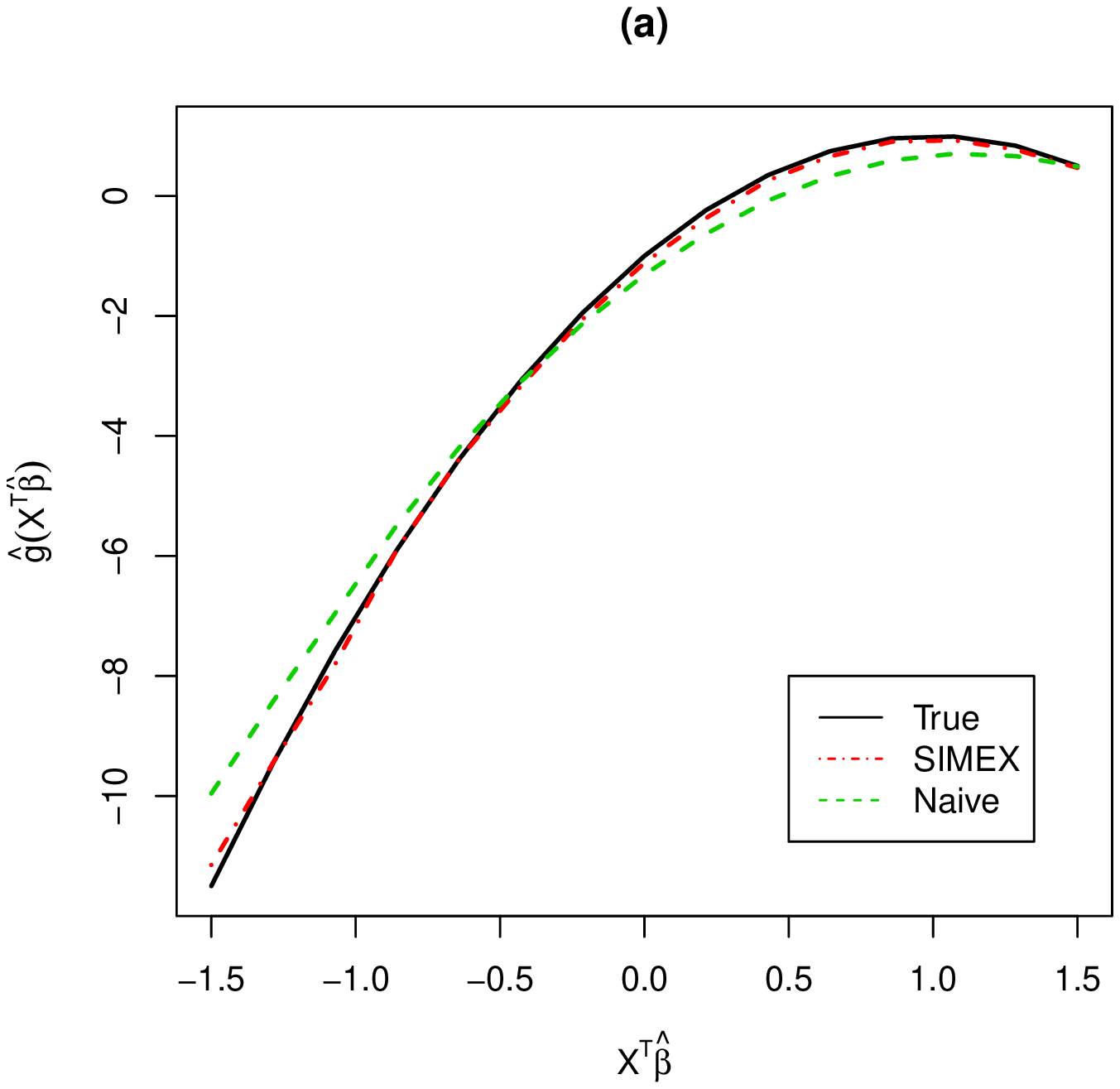}
    \includegraphics[width=6cm,height=6cm]{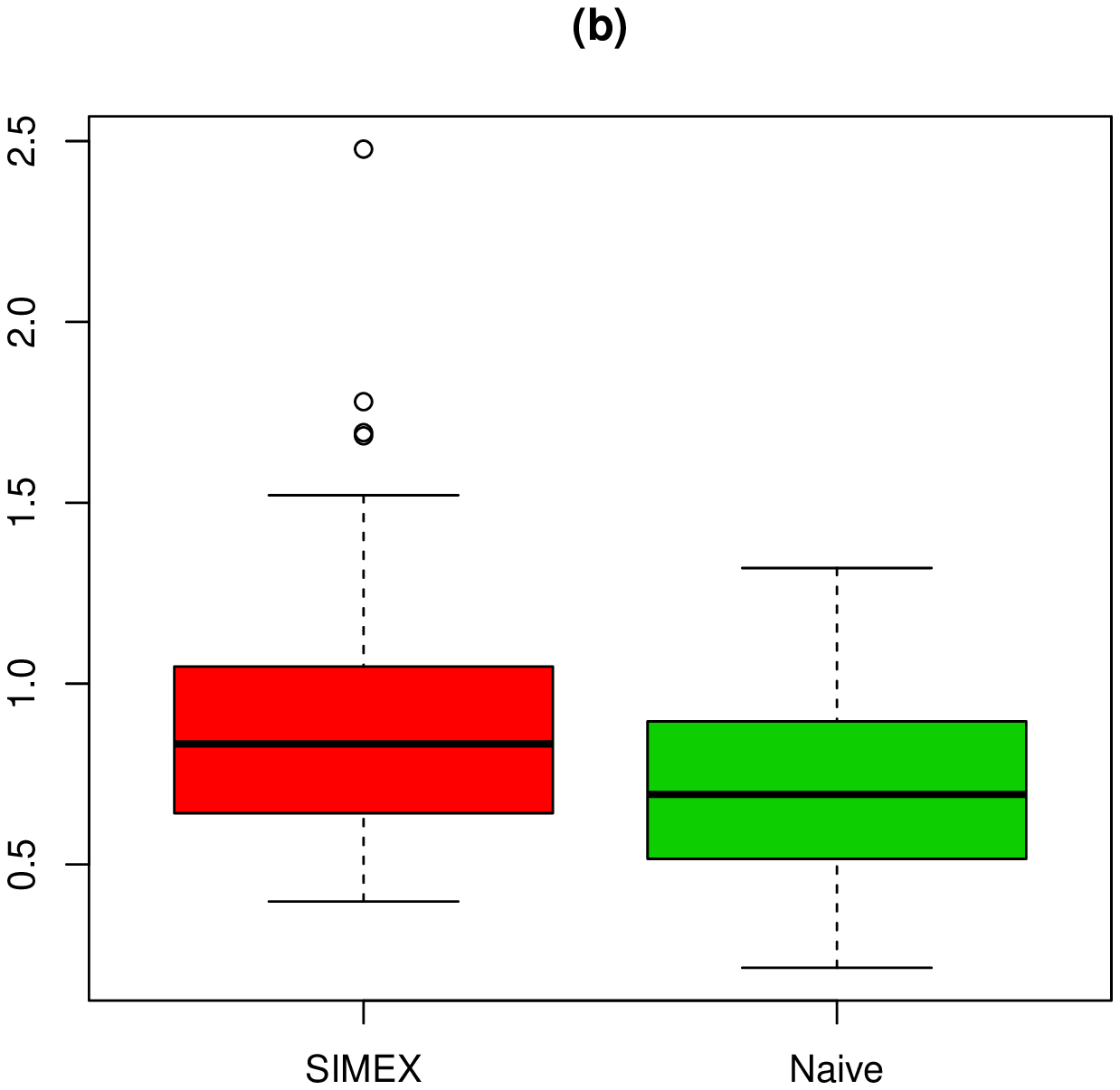}
\caption{{ \it (a) The real curve (solid curve), the naive estimated curve (dashed curve) and the SIMEX  estimated curve (dotted-dashed curve) for the link function $g(t)$ when $n=100$ and $\sigma_u=0.4$. (b) The boxplots of the 500 RMSE values for the estimate of $g(t)$. }}
   \label{fighatg}
\end{figure}

\subsection{Real data analysis}

We now analyze a data set from the Framingham Heart
Study to illustrate the proposed method. The data set contains 5 variables with 1615 males and it  has been used
by many authors to illustrate semiparametric partially linear models  (see  \citeasnoun{LH1999}, \citeasnoun{WBC2011}).
We are interested in whether the age and the serum cholestoral have an effect to the blood pressure. We use the proposed model to analyze the Framingham data to compare the SIMEX and naive estimators. We use the Epanechnikov kernel and the bandwidths $h=0.0589$ and $h_1=h_2=0.2309$.
Let $Y$ be their average blood pressure in a fixed
two-year period, $W_1$ and $W_2$ be  the standardized
variable for the logarithm of the serum cholestoral level ($\log(SC)$) and age, respectively.
Similar to \citeasnoun{LH1999}, $W_1$  is subject to the measurement error $U$ and $\sigma_u^2$
is estimated to be 0.2632 by two replicates experiments. Figure \ref{figSC} shows
the duplicated  serum cholestoral level measurements from 1615 males. The estimators of $\beta$
and $g(\cdot)$ based on the SIMEX and naive methods are reported in Table \ref{hp}, Figure \ref{figpr} and Figure \ref{fignpr}.

\begin{figure}[htbp]
    \centering
    \includegraphics[width=6cm,height=6cm]{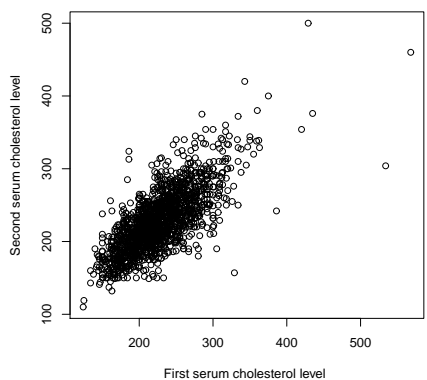}
\caption{{ \it Duplicated serum cholestoral level measurements from 1615 males in Framingham Heart
Study.  }}
   \label{figSC}
\end{figure}

\begin{figure}[htbp]
    \centering
    \includegraphics[width=6cm,height=6cm]{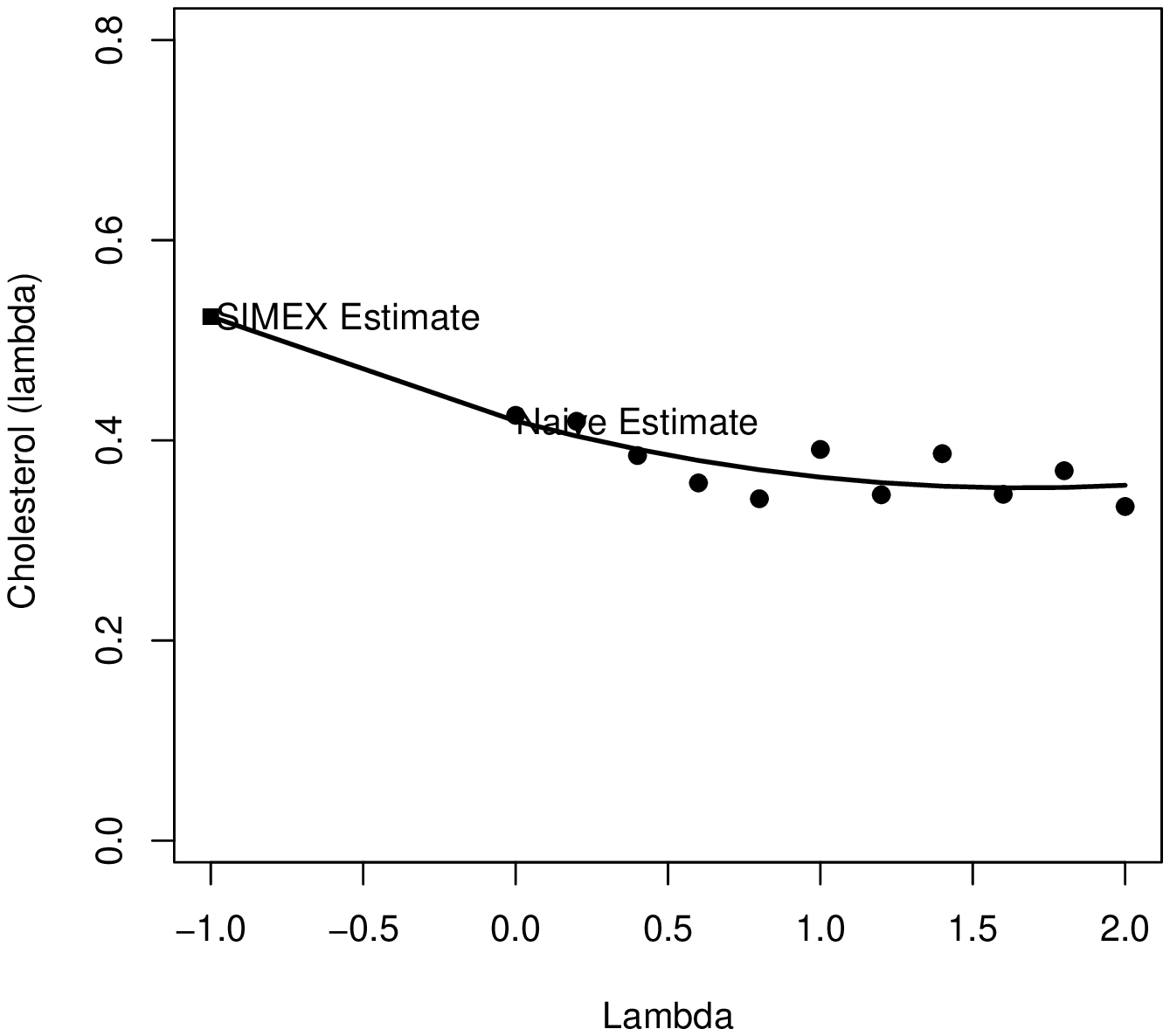}
    \includegraphics[width=6cm,height=6cm]{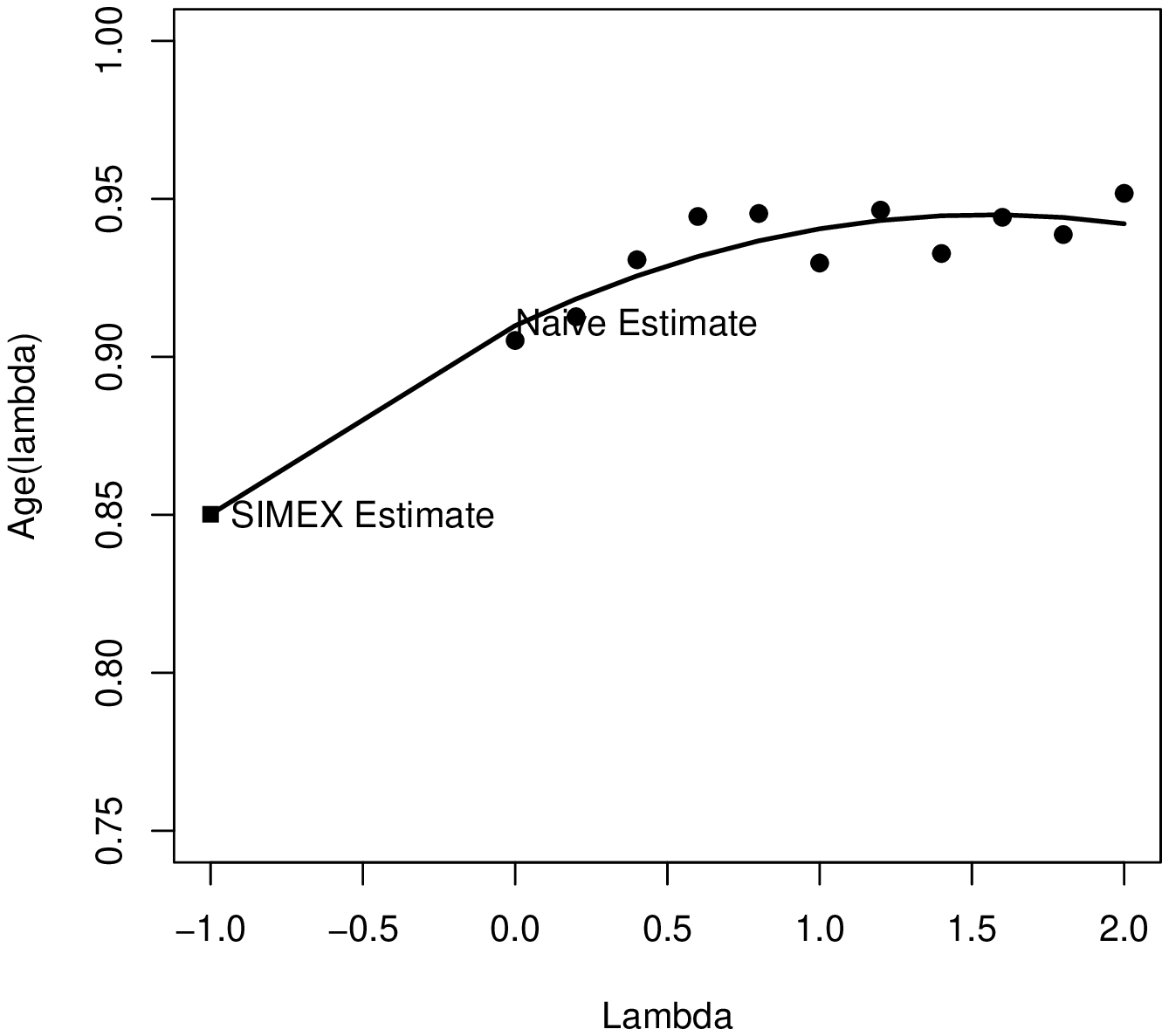}
\caption{{ \it The extrapolated point estimators for the Framingham data. The simulated estimates $\{\hat{\beta}(\lambda),\lambda\}$ are plotted (dots), and the fitted quadratic function (solid lines) is extrapolated to $\lambda=-1$. The extrapolation results are the SIMEX estimates (squares).  }}
   \label{figpr}
\end{figure}

\begin{table}[h]
  \caption{\label{hp} {\rm
 The estimators of the parameters
obtained by the SIMEX and naive methods for the Framingham data. }}
\vskip 8pt
\centering \tabcolsep 1.40cm
\begin{tabular}{ ccc }
\hline
Method &$\log(SC)$& Age\\
 \hline
~~SIMEX~ &$0.5237$&$0.8502$\\
~~Naive~ &$0.4194$&$0.9099$  \\\hline
\end{tabular}%
\end{table}

\begin{figure}[htbp]
    \centering
    \includegraphics[width=9cm,height=6cm]{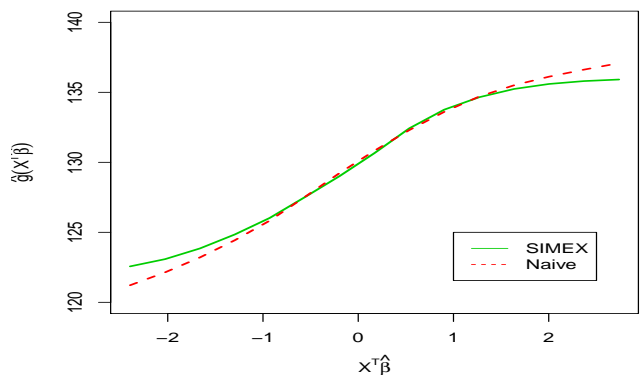}
\caption{{ \it The link function estimators for the Framingham data: the naive estimated curve (solid curve) and the SIMEX estimated curve (dashed curve).  }}
   \label{fignpr}
\end{figure}

From Table \ref{hp}, we can see that the SIMEX estimate of the index coefficient $\log (SC)$ is larger, while the SIMEX estimate of Age is smaller than the naive estimate. The results also show that the serum cholestoral and the age
are statistically significant. Figure \ref{figpr} shows  the trace of the extrapolation step for the SIMEX algorithm. The estimates of the two index coefficients for the different $\lambda$ values are plotted. The SIMEX estimates of index coefficients correspond to $-1$ on the horizontal axis, while the naive estimates correspond to 0 on the horizontal axis.
Figure \ref{fignpr} shows that the estimates of $g(\cdot)$  are obtained by the SIMEX method and the naive method.
The patterns of the two curves are similar. Table \ref{hp} and Figure \ref{fignpr} show that  the age and the serum cholestoral have a  positive
association with the blood pressure.
As expected, when the  measurement error is taken into account, we find a somewhat stronger positive
association between the serum cholestoral and the blood pressure. \citeasnoun{LH1999} also analyzed the relationship among the blood pressure, the age,
and the logarithm of serum cholesterol level by the partially linear errors-in-variables model, where the logarithm of serum cholesterol level was the covariate of the corresponding parameter and the age was a scalar covariate of the corresponding unknown function. When they accounted for the measurement error, the estimator of the parameter was larger than that of ignoring the measurement error. It implied that the blood pressure and
the serum cholestoral had  a stronger positive correlation when considering the measurement error. The estimator of the unknown function showed that
the age was positively associated with the blood pressure. Our findings basically agree with those discovered in \citeasnoun{LH1999}.

\section{Conclusion}

We propose the SIMEX estimation of the index parameter and the unknown link function for single-index models with covariate measurement error.
The asymptotic normality of the estimator of the index parameter and the asymptotic bias and variance of the estimator of the unknown link function are derived under some regularity conditions. The proposed index parameter estimator is root-$n$ consistent, which is similar to that of
the estimator of a parameter without measurement error, but the asymptotic covariance has a complicated  form. The asymptotic variance of the estimator of the unknown link function is of order $(nh_2)^{-1}$. Our simulation studies indicate that the proposed method works well in practice.

The proposed method can be extended to some other models, including partially linear single-index models with measurement error in nonparametric components and generalized single-index models with covariate measurement error. We can also extend to single-index measurement error models with cluster data by assuming working independence in the estimating equations. Future study is needed to investigate how to take into account the within-cluster correlation for cluster data to improve the
efficiency of the estimator of the index parameter for single-index measurement error models with cluster data.



\vskip 24pt
\section*{Appendix}
\renewcommand{\theequation}{A.\arabic{equation}}
\setcounter{equation}{0}

The following notation will be used in the proofs of the lemmas and theorems. Set $\beta_0$ be true value, $\mathcal{B}_n=\{\beta: \|\beta\|=1, \|\beta-\beta_0\|\leq c_1 n^{-1/2}\}$ for some positive constant $c_1$. Let
$f_\lambda(\cdot)$ be the density function of $\beta^TW_b(\lambda)$. Note that if $\lambda=0$, $f_0(\cdot)$ is the density function of $\beta^TW$.

\begin{lemm}\label{CR}
Let $(\zeta_1, \eta_1),\ldots, (\zeta_n, \eta_n)$ be i.i.d. random vectors, where $\eta_i$'s are scalar random variables. Assume further that $E|\eta_1|^s<\infty$, and $\sup_x\int|y|^sf(x,y)dy<\infty$, where $f(\cdot,\cdot)$ denotes the joint density of $(\zeta_1,\eta_1)$. Let $K(\cdot)$ be a bounded positive function with a bounded support, satisfying a Lipschitz condition. Then
$$\sup\limits_x\Big|{1\over n}\sum\limits_{i=1}^n\left\{K_h(\zeta_i-x)\eta_i-E[K_h(\zeta_i-x)\eta_i]\right\}\Big|=O_p\left(\left\{{\log(1/h)\over nh}\right\}^{1/2}\right),$$
provided that $n^{2\epsilon-1}h\rightarrow\infty$ for some $\epsilon<1-s^{-1}$.
\end{lemm}

\textbf{Proof:} This follows immediately from the result that was obtained by \citeasnoun{MS1982}.

\begin{lemm}\label{CGR}
Suppose that conditions (C1)--(C4) hold. Then
$$\sup\limits_{t\in\mathcal{T},\beta\in \mathcal{B}_n}\big|\hat{g}(\beta,\lambda;t)-g(\lambda;t)\big|=O_p\big((nh/\log n)^{-1/2}+h^2\big)$$
and
$$\sup\limits_{t\in\mathcal{T},\beta\in \mathcal{B}_n}\big|\hat{g}'(\beta,\lambda;t)-g'(\lambda;t)\big|=O_p\big((nh^3/\log n)^{-1/2}+h\big).$$
\end{lemm}

\textbf{Proof:} By the theory of least squares, we have
\begin{eqnarray}\label{HAG}
(\hat{g}(\beta,\lambda;t),h\hat{g}'(\beta,\lambda;t))^T=S_n^{-1}(\beta,\lambda;t)\xi_n(\beta,\lambda;t),
\end{eqnarray}
where $$S_n(\beta,\lambda;t)=\left(\begin{array}{cc}S_{n,0}(\beta,\lambda;t)&h^{-1}S_{n,1}(\beta,\lambda;t)\\
h^{-1}S_{n,1}(\beta,\lambda;t)&h^{-2}S_{n,2}(\beta,\lambda;t)
\end{array}\right)$$
and
$$\xi_n(\beta,\lambda;t)=(\xi_{n,0}(\beta,\lambda;t)),\xi_{n,1}(\beta,\lambda;t))^T$$
with
$$\xi_{n,l}(\beta,\lambda;t)={1\over n}\sum\limits_{i=1}^nY_i\left({\beta^TW_{ib}(\lambda)-t}\over h\right)^lK_h(\beta^TW_{ib}(\lambda)-t)$$
for $l=0,1,2.$ A simple
calculation yields, for $l=0,1,2,3,$
\begin{eqnarray}\label{ESN}E[h^{-1}S_{n,l}(\beta,\lambda;t)]=f_\lambda(t)\mu_l+O(h).\end{eqnarray}
By Lemma \ref{CR}, we have
$$h^{-1}S_{n,l}(\beta,\lambda;t)-E[h^{-1}S_{n,l}(\beta,\lambda;t)]=O_p\left(\left\{{\log(1/h)\over nh}\right\}^{1/2}\right),$$
which, combining with (\ref{ESN}), proves that, for $t\in\mathcal{T}$ and $\beta\in \mathcal{B}_n$,
\begin{eqnarray}\label{SN}
h^{-1}S_{n,l}(\beta,\lambda;t)=f_\lambda(t)\mu_l+O_p\left(\left\{{\log(1/h)\over nh}\right\}^{1/2}+h\right), ~~~l=0,1,2,3.
\end{eqnarray}
It can be obtained immediately that
$$S_n(\beta,\lambda;t)=S(\lambda;t)+O_p\left(\left\{{\log(1/h)\over nh}\right\}^{1/2}+h\right),$$
where $S(\lambda;t)=f_\lambda(t)\otimes {\rm diag}(1,\mu_2)$, and $\otimes$ is the Kronecker product.

Denote
$$\xi_{n,l}^*(\beta,\lambda;t)={1\over n}\sum\limits_{i=1}^n[Y_i-g(\lambda;\beta^TW_{ib}(\lambda))]\left({\beta^TW_{ib}(\lambda)-t}\over h\right)^lK_h(\beta^TW_{ib}(\lambda)-t)$$ and
$$\xi_n^*(\beta,\lambda;t)=\Big(\xi_{n,0}^*(\beta,\lambda;t), \xi_{n,1}^*(\beta,\lambda;t)\Big)^T.$$
Note that
\begin{eqnarray}\label{EXi}E(\xi_n^*(\beta,\lambda;t))=O(n^{-1/2}).\end{eqnarray}
By Lemma 1 and (\ref{EXi}), it can be shown that
\begin{eqnarray}\label{XIN}\xi_n^*(\beta,\lambda;t)=O_p\left(\left\{{\log(1/h)\over nh}\right\}^{1/2}+n^{-1/2}\right).
\end{eqnarray}
By applying Taylor's expansion for $g(\lambda;\beta^TW_{ib}(\lambda))$ at $t$, we can prove that
\begin{eqnarray*}\xi_{n,0}(\beta,\lambda;t)-\xi_{n,0}^*(\beta,\lambda;t)&=&S_{n,0}(\beta,\lambda;t)g(\lambda;t)+S_{n,1}(\beta,\lambda;t)hg'(\lambda;t)\\
&&~+{1\over 2}h^2S_{n,2}(\beta,\lambda;t)g''(\lambda;t)+o_p\{h^2+(nh)^{-1/2}\}
\end{eqnarray*}
and
\begin{eqnarray*}\xi_{n,1}(\beta,\lambda;t)-\xi_{n,1}^*(\beta,\lambda;t)&=&S_{n,1}(\beta,\lambda;t)g(\lambda;t)+S_{n,2}(\beta,\lambda;t)hg'(\lambda;t)\\
&&~+{1\over 2}h^2S_{n,3}(\beta,\lambda;t)g''(\lambda;t)+o_p\{h^2+(nh)^{-1/2}\}
\end{eqnarray*}
uniformly hold
in $t\in\mathcal{T}$ and $\beta\in \mathcal{B}_n$. Hence
\begin{eqnarray*}\xi_{n}(\beta,\lambda;t)-\xi_{n}^*(\beta,\lambda;t)&=&S_{n}(\beta,\lambda;t)
\left(\begin{array}{c}g(\lambda;t)\\
h g'(\lambda;t)
\end{array}\right)+{1\over2}h^2\left(\begin{array}{c}S_{n,2}(\beta,\lambda;t)g''(\lambda;t)\\
S_{n,3}(\beta,\lambda;t)g''(\lambda;t)
\end{array}\right)\\
&&~+o_p\{h^2+(nh)^{-1/2}\}.
\end{eqnarray*}
Combining this with (\ref{HAG})--(\ref{SN}) yields
\begin{eqnarray}\label{gg1}
\left(\begin{array}{c}\hat{g}(\lambda;t)-g(\lambda;t)\\\nonumber
h \{\hat{g}'(\lambda;t)-g'(\lambda;t)\}
\end{array}\right)&=&S^{-1}(\lambda;t)\xi_n^*((\beta,\lambda;t))\\
&&+{1\over2}h^2\left(\begin{array}{c}\mu_{2}g''(\lambda;t)\\
{\mu_{3}\over \mu_2}g''(\lambda;t)
\end{array}\right)+o_p(h^2+(nh)^{-1/2}).
\end{eqnarray}
This together with (\ref{XIN}) proves Lemma \ref{CGR}.

\textbf{Proof of Theorem \ref{AN}:} Assume $\beta(\lambda)$ is the true value based on the model $E[Y|\beta^T(\lambda) W_b(\lambda)]=g(\beta^T(\lambda) W_b(\lambda))$. Using Lemma  \ref{CGR} and  the similar method in Theorem 1 of \citeasnoun{CXZ2010}, we have
\begin{eqnarray*}
\sqrt{n}\Big(\hat{\beta}_b(\lambda)-\beta(\lambda)\Big)=\sqrt{n}J_{\beta^{(r)}(\lambda)}A_n^{-1}(\beta(\lambda),
\lambda)B_n(\beta(\lambda),\lambda)+o_p(1),
\end{eqnarray*}
where
\begin{eqnarray*}
A_n(\beta(\lambda),\lambda)={1\over n}\sum\limits_{i=1}^n\Big[g'\Big(\lambda; \beta^T(\lambda) W_{ib}(\lambda)\Big)\Big]^2J_{\beta^{(r)}(\lambda)}^T{\widetilde{W}}_{ib}(\lambda){\widetilde{W}}^T_{ib}(\lambda)J_{\beta^{(r)}(\lambda)}
\end{eqnarray*}
and
\begin{eqnarray*}
B_n(\beta(\lambda),\lambda)={1\over n}\sum\limits_{i=1}^n\epsilon_{ib}(\lambda)g'\Big(\lambda; \beta^T(\lambda) W_{ib}(\lambda)\Big)J_{\beta^{(r)}(\lambda)}^T{\widetilde{W}}_{ib}(\lambda)
\end{eqnarray*}
with $\epsilon_{ib}(\lambda)=Y_i-g\Big(\lambda; \beta^T(\lambda) W_{ib}(\lambda)\Big)$.

Extrapolation step deduces that
\begin{eqnarray}\label{betala}
\sqrt{n}\Big(\hat{\beta}(\lambda)-\beta(\lambda)\Big)=J_{\beta^{(r)}(\lambda)}\mathcal{A}^{-1}(\beta(\lambda),\lambda)n^{-{1\over 2}}\sum\limits_{i=1}^n\eta_{iB}(\beta(\lambda),\lambda)+o_p(1),
\end{eqnarray}
where $\eta_{iB}(\beta(\lambda),\lambda)=\displaystyle {1\over B}\sum\limits_{b=1}^B\epsilon_{ib}(\lambda)g'\Big(\lambda; \beta^T(\lambda) W_{ib}(\lambda)\Big)J_{\beta^{(r)}(\lambda)}^T{\widetilde{W}}_{ib}(\lambda).$

Then, using (\ref{betala}), the limit distribution of $\sqrt{n}\Big(\hat{\beta}(\Lambda)-\beta(\Lambda)\Big)$ is multivariate normal
distribution with mean zero and covariance $\Sigma$.

$\hat{\mathbf{\Gamma}}$ in the extrapolation step is obtained by minimizing $\{{\rm Res}(\mathbf{\Gamma})\}\{{\rm Res}(\mathbf{\Gamma})\}^T$. The estimating equation for $\hat{\mathbf{\Gamma}}$ is $0=s(\mathbf{\Gamma}){\rm Res}(\mathbf{\Gamma})$, where $s^T(\mathbf{\Gamma})=\{\partial/\partial (\mathbf{\Gamma})^T\}{\rm Res}(\mathbf{\Gamma})$. Then, we have
$$\sqrt{n}(\hat{\mathbf{\Gamma}}-\mathbf{\Gamma})\stackrel{\mathcal {L}}{\longrightarrow}N\{0,\Sigma(\mathbf{\Gamma})\}.$$

Because $\hat{\beta}_{\rm SIMEX}=\mathcal{G}(-1,\hat{\mathbf{\Gamma}})$, the SIMEX estimator is asymptotically normal with asymptotic variance
$$
\mathcal{G}_{\mathbf{\Gamma}}(-1,\mathbf{\Gamma})\Sigma(\mathbf{\Gamma})\{\mathcal{G}_{\mathbf{\Gamma}}(-1,\mathbf{\Gamma})\}^T.
$$

\textbf{Proof of Theorem \ref{ANG}:} Note that $\|\hat{\beta}_{\rm SIMEX}-\beta\|=O_p(n^{-1/2})$, similar to the proof of  (\ref{gg1}), we have
\begin{eqnarray}\nonumber
&&\hat{g}_b(\lambda; t_0)-g(\lambda;t_0)-{1\over 2}h_2^2\mu_2g''(\lambda; t_0)\\\label{hatgb}
&=&[f_\lambda(t_0)]^{-1}\displaystyle {1\over n}\sum\limits_{i=1}^n\left\{[Y_i-g(\lambda;\beta^TW_{ib}(\lambda))]K_{h_2}(\beta^TW_{ib}(\lambda)-t_0)\right\}\\\nonumber
&&+o_p\{h_2^2+(nh_2)^{-1/2}\}.
\end{eqnarray}
Using (\ref{hatgb}) and the decomposition of \citeasnoun{CLKS1996}, since $B$ is fixed and $\hat{g}(\lambda; t_0)
=B^{-1}\sum\limits_{b=1}^B hat{g}_b(\lambda; t_0)$, we have
\begin{eqnarray}\nonumber
&&\hat{g}(\lambda; t_0)-g(\lambda;t_0)-{1\over 2}h_2^2\mu_2g''(\lambda; t_0)\\\label{hatgll}
&=&[f_\lambda(t_0)]^{-1}\displaystyle {1\over n}\sum\limits_{i=1}^n\left\{B^{-1}\sum\limits_{b=1}^B[Y_i-g(\lambda;\beta^TW_{ib}(\lambda))]K_{h_2}(\beta^TW_{ib}(\lambda)-t_0)\right\}\\\nonumber
&&+o_p\{h_2^2+(nh_2)^{-1/2}\}.
\end{eqnarray}

 If $\lambda=0$, (\ref{hatgll}) becomes
 \begin{eqnarray*}
&&\hat{g}(0; t_0)-g(0;t_0)-{1\over 2}h_2^2\mu_2g''(0; t_0)\\
&=&[nf_0(t_0)]^{-1}\displaystyle {1\over n}\sum\limits_{i=1}^n[Y_i-g(0;\beta^TW_i)]K_{h_2}(\beta^TW_{i}-t_0)
+o_p\{h_2^2+(nh_2)^{-1/2}\},
\end{eqnarray*}
which has mean zero and the following asymptotic variance
\begin{eqnarray}\label{vairance0}
[nh_2f_0(t_0)]^{-1}{\rm var}(Y|\beta^TW=t_0)\nu_2.
\end{eqnarray}

For $\lambda>0$, using the similar argument of (A8) in \citeasnoun{CMR1999}, we have
$${\rm var}(\hat{g}(\lambda; t_0))=O\{(nh_2B)^{-1}\}+O(n^{-1}),$$
while for $\lambda=0$,
$${\rm var}(\hat{g}(\lambda; t_0))=O\{(nh_2)^{-1}\}.$$
Then, for $B$ sufficiently large, the variability of $\hat{g}(\lambda; \cdot)$ is negligible for $\lambda>0$ compared to $\lambda=0$. Hence, in what follows, we will ignore this variability by treating $B$ as if it was equal to infinity.

We obtain $\hat{\mathbb{A}}$ by solving the following equation
\begin{eqnarray}\label{GEQ}0=
\sum\limits_{\lambda\in\Lambda}\{\hat{g}(\lambda;t_0)-\mathcal{G}(\lambda,\mathbb{A})\}\gamma(\lambda,\mathbb{A}).\end{eqnarray}
Applying the Taylor expansion for the left side of (\ref{GEQ}), we obtain
$$0=\sum\limits_{\lambda\in\Lambda}\{\hat{g}(\lambda;t_0)-\mathcal{G}(\lambda,\mathbb{A})\}\gamma(\lambda,\mathbb{A})-
\sum\limits_{\lambda\in\Lambda}\gamma(\lambda,\mathbb{A})\gamma^T(\lambda,\mathbb{A})(\hat{\mathbb{A}}-\mathbb{A}),
$$
Hence,
\begin{eqnarray}\label{hataa}
\hat{\mathbb{A}}-\mathbb{A}=\left\{\sum\limits_{\lambda\in\Lambda}\gamma(\lambda,\mathbb{A})\gamma^T(\lambda,\mathbb{A})\right\}^{-1}
\sum\limits_{\lambda\in\Lambda}\{\hat{g}(\lambda;t_0)-\mathcal{G}(\lambda,\mathbb{A})\}\gamma(\lambda,\mathbb{A}).
\end{eqnarray}
The left side of (\ref{hataa}) has approximate mean
 \begin{eqnarray*}
\left\{\sum\limits_{\lambda\in\Lambda}\gamma(\lambda,\mathbb{A})\gamma^T(\lambda,\mathbb{A})\right\}^{-1}
\sum\limits_{\lambda\in\Lambda}{1\over 2}h_2^2 \mu_2g''(\lambda;t_0)\gamma(\lambda,\mathbb{A}),
\end{eqnarray*}
and its approximate variance is given by
$$[nh_2f_0(t_0)]^{-1}\nu_2 {\rm var}(Y|\beta^TW=t_0)\left\{\sum\limits_{\lambda\in\Lambda}\gamma(\lambda,\mathbb{A})\gamma^T(\lambda,\mathbb{A})\right\}^{-1}D
\left\{\sum\limits_{\lambda\in\Lambda}\gamma(\lambda,\mathbb{A})\gamma^T(\lambda,\mathbb{A})\right\}^{-1}.$$
Because $\hat{g}_{\rm SIMEX}(t_0)=\mathcal{G}(-1,\hat{\mathbb{A}})$, so that its
 asymptotic bias is
$$C(\Lambda,\mathbb{A})\sum\limits_{\lambda\in\Lambda}{1\over 2}h_2^2 \mu_2g''(\lambda;t_0)\gamma(\lambda,\mathbb{A}),$$
and its  asymptotic variance is
$$[nh_2f_0(t_0)]^{-1}\nu_2 {\rm var}(Y|\beta^TW=t_0)C(\Lambda,\mathbb{A})DC^T(\Lambda,\mathbb{A}).$$
This completes the proof.

\bibliographystyle{dcu}
\bibliography{reference}

\end{document}